\documentclass[preprint2]{aastex}

\usepackage{natbib}
\shorttitle{CME remote and in-situ observations}
\shortauthors{M\"ostl et al.}

\begin{document}


\title{Linking remote imagery of a coronal mass ejection to its in situ signatures at 1 AU}


\author{C. M\"ostl\altaffilmark{1,2} and C. J. Farrugia\altaffilmark{3} }

\author{M. Temmer\altaffilmark{2}, C. Miklenic\altaffilmark{1,2}, and A. M. Veronig\altaffilmark{2} }

\author{A. B. Galvin\altaffilmark{3}, M. Leitner\altaffilmark{4} and H. K. Biernat\altaffilmark{1,2} }

\affil{Space Research Institute, Austrian Academy of Sciences, Graz A-8042,
Austria}
\affil{Institute of Physics, University of Graz, A-8010, Austria}
\affil{Space Science Center and Dept. of Physics, University of New
Hampshire, Durham, NH 03824, USA }
\affil{Institute for Astro- and Particle Physics, University of Innsbruck, A-6020, Austria}
\email{christian.moestl@oeaw.ac.at}




\begin{abstract}

In a case study (June 6-7, 2008) we report on how the internal structure of a coronal mass ejection (CME)
 at 1 AU can be anticipated from remote observations of white-light images of the heliosphere.
 Favorable circumstances are the absence of fast equatorial solar wind streams and a low CME velocity
  which allow us to relate the imaging and in-situ data in a straightforward way.
The STEREO-B spacecraft encountered typical signatures of a magnetic flux rope inside an interplanetary CME (ICME)
whose axis was inclined at 45$^{\circ}$ to the solar equatorial plane. Various CME direction--finding
techniques yield consistent results to within $15^{\circ}$.
 Further, remote images from  STEREO-A
 show that (1) the CME is unambiguously connected to the ICME 
 and can be tracked all the way to 1 AU,  (2) the particular arc-like morphology of the CME points to an inclined axis,
 and (3) the three--part structure of the CME may be plausibly related to the in situ data.
This is a first step in predicting both the direction of travel and the internal structure of CMEs
from complete remote observations between the Sun and 1 AU, which is one of the main requirements for forecasting the geo-effectiveness of CMEs.

\end{abstract}





\section{Introduction}

Coronal mass ejections (CMEs) are violent expulsions of plasma and magnetic flux
from the outer solar atmosphere. 
Understanding their propagation characteristics and their internal structure is one of the main goals of the NASA twin-spacecraft
 \emph{Solar Terrestrial Relations Observatory} \citep[STEREO, ][]{kai08}.
Recently, there have been many efforts to obtain the
direction of propagation of CMEs both close to the Sun and in interplanetary space \citep[e.g.][]{mie08, the09, tem09, dav09b}.
Using interplanetary observations by the STEREO Heliospheric Imagers \citep[HIs,][]{eyl09,har09},
\cite{dav09} were able to predict the CME direction of propagation and arrival time at Earth with good accuracy. However, a major
issue in forecasting the geo-effects of a given CME is its orientation and its internal magnetic field since
long-duration southward fields lead to strong geomagnetic activity
\citep[e.g.][]{far93c,zha98}.
Thus the above-mentioned techniques need to be compared with the magnetic field and plasma signatures of CMEs as measured in situ
in the solar wind (interplanetary coronal mass ejections, ICMEs). Those ICMEs
containing a rotating magnetic field vector are particularly suited for these comparisons. These
are generally called magnetic flux ropes (MFRs), and may be  magnetic clouds (MCs) if  they satisfy other criteria, namely, above-average magnetic field strengths ${\bf B}$,
low proton temperature ($T_p$) and beta ($\beta_p$), and a large rotation of ${\bf B}$ \citep{bur81}.

Using SOHO/LASCO coronagraph images, \cite{cre04} have shown that CMEs  have a different morphology seen along or perpendicular to the axis of symmetry.
In this Letter we relate, for the first time, both the morphology and direction of a CME seen with the STEREO-HI
instrument imaging the heliosphere to properties of the associated ICME measured in-situ.
Our study thus extends previous efforts to predict the magnetic field orientation in MFRs from solar disk and coronagraph observations \citep[e.g.][]{bot98,yur01,yur08}.
The big advantage is that the CME can be seen in the HI images from a distant spacecraft (STEREO-Ahead) passing over the other spacecraft (STEREO-Behind) at 1~AU, thus bridging the gap between remote and in-situ observations which existed before the STEREO era.

The CME in question left the Sun $\sim$ 21:00 UT, June 1, 2008 and its front boundary arrived at STEREO-B at 22:39 UT on June 6, 2008.
In previous work \cite{rob09} used coronagraph and on-disk extreme ultraviolet (EUV) images
to show that this CME was not accompanied by typical on-disk signatures (filament eruption, flare, dimming, EUV wave),
the main reason being that the CME was of the streamer-blowout type lifting off slowly from high in the corona
 ($R_{\odot} \approx 1.15-1.4 $). This previously unrecognized type of CME is a good candidate for causing so-called ``problem storms",
i.e., those without an obvious solar origin \cite[e.g.][]{zha07}. In this Letter, we complement and extend the study
 of \cite{rob09} by modeling STEREO-B observations to obtain a more complete view of the MFR
inside the ICME at 1 AU and discuss how its structure and direction could have been forecast from remote observations.
CME-ICME events suitable for this kind of study are very rare during this unusually quiet solar minimum \citep[see][]{dav09, moe09b}. This event is extraordinarily well suited for this kind of analysis because (i) the absence of equatorial coronal holes (and thus fast solar wind streams) allows a straightforward interpretation of the HI images \citep{lug08}, and (ii) it is a slow CME, justifying partly an assumption of constant velocity inherent to one of the applied techniques.

\section{In situ observations of the ICME at STEREO-B}

At 00:00 UT, June 7 2008, STEREO-B was $25.38^{\circ}$ east of Earth at 1.0545~AU and STEREO-A $29.26^{\circ}$ west at 0.9567~AU,
with a separation of $54.64^{\circ}$. Fig.~\ref{fig:data} plots STEREO-B magnetic field data in
Radial-Tangential-Normal (RTN) coordinates and plasma bulk parameters at 1 min resolution, from the IMPACT/MAG (\cite{acu08,luh08})
and PLASTIC \citep{gal08} instruments, respectively. STEREO-B encountered a MFR between 22:39 UT, June 6 and 12:27 UT,
June 7, 2008 (inner two vertical solid lines).
 This interval is partly an outcome of our reconstruction technique \citep[see][]{hu04} and partly determined by eye so that it encompasses the  smoothly rotating {\bf B}-vector.


The two outer solid lines indicate a forward ($t_s$=June 6 15:35 UT) and a very weak reverse shock $t_r$= June 7 20:48 UT, see e.g. \cite{gos98}.
 There are two clear proton enhancements of $N_p \approx 30$ cm$^{-3}$ on either side of the MFR, the first one being,
  in part, the sheath behind the forward shock.  The front of the second density
  peak arrives at $t_c \approx$ 12:00 UT, June 7. For most of the MFR, $T_p$, is higher than that expected for normal solar wind
  expansion (red trace, \cite{lop87}). This precludes the MFR from being a magnetic cloud.

We modeled these data using Grad-Shafranov reconstruction \citep{hu02} and force-free fitting \citep[the latter shown in parentheses,][]{lep90}. (For recent multi-spacecraft validation of the GS method see \cite{liu08,moe09b,moe09}.) The MFR is right-handed with an axis orientation of $\theta=51(37)^{\circ} $, $\varphi=278(326)^{\circ}$ in RTN coordinates
($\theta$ is the inclination to the RT plane, $\varphi$ is measured from R ($0^{\circ}$) towards T ($90^{\circ}$)). Thus the MFR axis points roughly north-east as shown in Fig. \ref{fig:3D}. The modeled axial field strength $B_0$=15.4 (20.3) nT, the radial scale size (diameter) $D=0.130 (0.155)$~AU, and the impact parameter (the closest  distance of the spacecraft to the MFR axis) is  $p=0.81 (0.60) \times D$. The toroidal and poloidal magnetic fluxes are $\Phi_t=0.72 (0.37) \times 10^{21} $~Mx and  $\Phi_p=1.19 (1.47) \times 10^{21} $~Mx/AU. These values are rather typical for ICMEs which are associated with medium-sized flares \citep{qiu07}.

Figure~\ref{fig:3D} shows a 3D plot of the local MFR reconstruction in the heliosphere, in a coordinate system centered on STEREO-B, seen from north looking down on the RT plane (close to the solar equatorial plane, top panel) and from another view point (bottom panel). Several field lines and its cross-section (color contour) are shown as well as its intersection (black contours) with the RT plane.  It emerges from the reconstruction that STEREO-B crossed the western flank of the MFR which intersected the RT plane  between  $24^{\circ}$ to $39^{\circ}$ east of Earth, while the MFR axis crosses at $33^{\circ}$. This is important when comparing the ICME structure at 1 AU to the results of the CME direction--finding techniques in Section 4. Further, in Fig.~\ref{fig:3D}
the spacecraft are indicated as STEREO-B (blue triangle), Wind (black diamond), STEREO-A (red X). The Thomson surface for STEREO-A is the blue sphere, the yellow sphere the size of the COR2 field of view (15 $R_{\odot}$). The CME directions from the method of triangulation (TR, yellow), forward modeling (FM, green), kinematic Fixed-$\Phi$ (KP, black) and elongation-fitting (EF, red) are indicated (see later sections and Table 1). The dashed (solid) black lines centered on STEREO-A indicate the boundaries of the HI1 (HI2) field of views.


\section{Remote imaging of the CME by STEREO-A}

Running difference images of the CME obtained from both cameras (HI1A: pointing $4-24^{\circ}$ away from Sun center, 40 min cadence; HI2A, $18.7-88.7^{\circ}$, 2  hours) of the Heliospheric Imagers aboard STEREO-A are shown in Fig. \ref{fig:hi}, from 15:29 UT, June 2, when the CME completely enters the HI1A FoV until 16:09 UT, June 6 ($\approx t_s$). The data have been put through the full reduction process, which includes corrections for flat-fielding, dark currents, background subtraction and the elimination of stars \citep{dav09b,bro09}. Only then can intriguing details of the morphology and evolution of the CME be seen. We strongly encourage the reader to look at the corrected running difference movies (see the electronic supplements), of which we show some selected images in Fig.~\ref{fig:hi}.

 In HI1A and HI2A, both the CME leading edge (LE) and core clearly show an arc-like shape, typical of a CME viewed  orthogonal  to its axis of symmetry \citep{cre04}, which is consistent with the derived MFR inclination of $\theta \approx 45^{\circ} \gg 0^{\circ}$. The intensity of the CME observed in the HI white-light images is determined by Thomson scattering of electrons \citep[e.g.][]{vou06}. Shortly after the CME enters the HI2A FoV it leaves the Thomson surface of strongest scattering (Fig.~\ref{fig:3D})
 and the intensity declines. The northern part vanishes earlier (at $\sim$50$^{\circ}$ elongation) than the southern part,
  which can be tracked all the way to STEREO-B. Assuming an evenly distributed density of the CME this effect is most likely caused by the CME and MFR inclination from which the angle $\varphi$ might be inferred: the northern part is tilted away from the observer, looking to the east (west) this yields $\varphi \approx 270 (90)^{\circ}$ for $\theta \approx +45^{\circ}$.
 This is again consistent with the MFR's moderate inclination ($\theta \approx 45^{\circ}$). The assumption made here that STEREO-B crosses the CME close to its apex or nose is supported by direction--finding techniques in section 4. In this case $\varphi$ is expected to be close to 270 or 90 degrees.  Additionally, there is a $180^{\circ}$ ambiguity in $\varphi$ depending on the sign of $\theta$; for a southward--pointing axial field, the opposite is obtained (see also conclusions).


\section{CME three--part structure and heliospheric propagation}

We now compare the  results of direction--finding results  applied close to the solar surface ($2-20 R_{\odot}$) with those used in the interplanetary medium ($20-220 R_{\odot}$),  and the ICME structure at 1 AU.

\begin{figure}[h]
\epsscale{.80}
\centerline{\includegraphics[width=20pc,  bb=0 0 300 500]{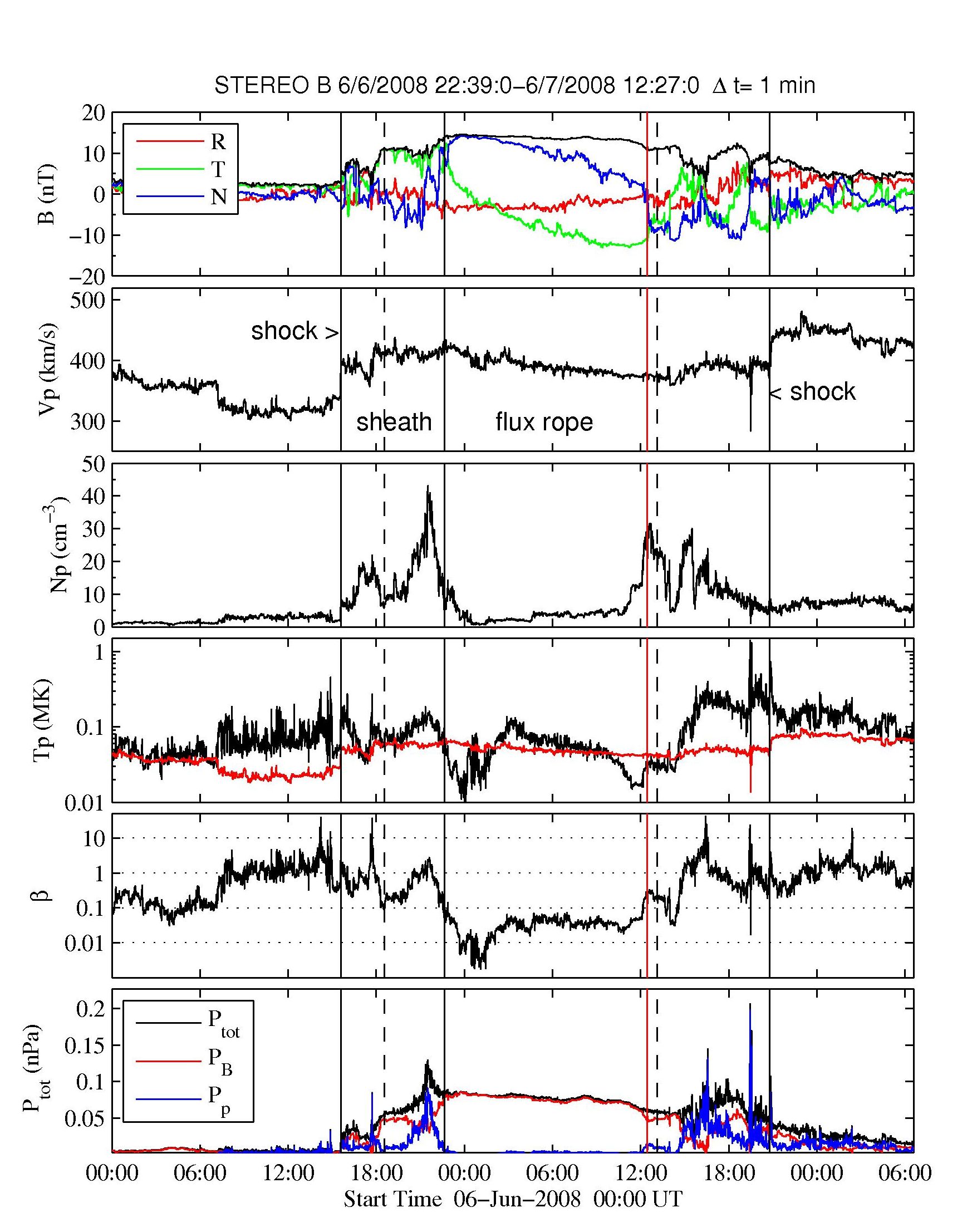}}
 \caption{Magnetic field and plasma data (STEREO-B). The interval between the two  middle solid lines is used for the reconstruction, the two outer solid lines indicate the forward and reverse shock. Dashed lines are the arrival times from the elongation fitting method for the CME leading edge (left) and  CME core (right). From top to bottom: magnetic field magnitude and magnetic field components in RTN coordinates (R pointing radially away from the Sun, T is the cross product of the solar rotation axis and R, N completes the right-handed triad),
  proton bulk velocity, proton number density,  proton temperature (black) and expected temperature (red), proton beta and the total, magnetic and plasma pressure.}
 \label{fig:data}
\end{figure}

First, the elongation $\epsilon(t)$ of the LE and core were manually measured in the HI FoVs and fitted with the \cite{she08}
formula \citep[e.g.][]{rou08,dav09b} to infer a constant velocity and direction (see Table~\ref{tab:cmedir}) as well as the arrival
times (18:34 UT, June 6 [LE] and 13:08 UT, June 7 [core]). The technique assumes that the CME is well represented by the features tracked in the white-light images which result from line-of-sight integration and Thomson scattering effects \citep{vou06}. The error bars for the direction and velocity follow from an error in $\epsilon(t)$ of $\pm 3^{\circ}$. The arrival times correspond very well with the arrival of the two density peaks on either side of the plasma void region (dashed vertical lines in Fig.~\ref{fig:data}). It thus becomes plausible how, in this case, the classic CME features seen in coronagraphs \citep{illing85} correspond to the ICME observations. The extended bright front, with the LE as its outermost edge, corresponds to the interval of the first density enhancement reaching up to the MFR; the void or dark cavity is the MFR;  and the core is the dense material trailing the MFR. We speculate that each double density peak bracketing the MFR
in Fig.~\ref{fig:data} arises from material originating in the corona (innermost density peaks adjacent to the MFR) and solar wind swept up by the CME in interplanetary space (outermost peaks), i.e., the sheaths \cite[see also][]{ril08}.

\begin{figure}[h]
\noindent\includegraphics[width=20pc,bb= 0 0 400 600]{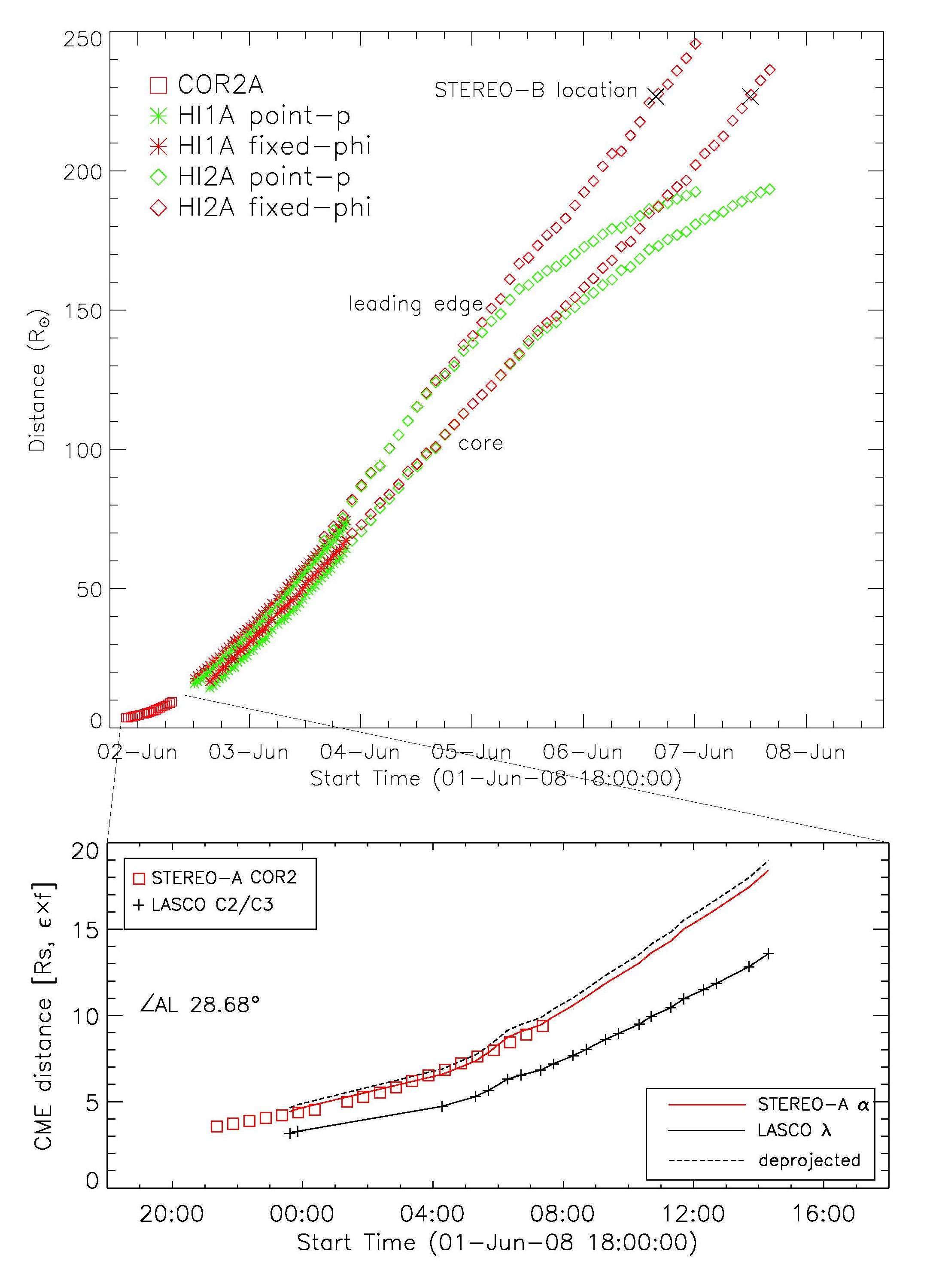}
 \caption{Time-distance plot of the CME leading edge and core. The elongation of the features in the HI images has been converted to
 distance by using the Point-P and Fixed-$\Phi$ methods ($\Phi=-30^{\circ}$ for the leading edge and $\Phi=-26^{\circ}$ for the core). The in-situ measured arrival times $t_s$ (shock) and $t_c$ (core) are indicated as '$\times$' at the position of STEREO-B (226~$R_\odot$). The inset shows the STEREO-A/COR2 and SOHO/LASCO observations used for the TR method. }
 \label{fig:cme1}
\end{figure}

Fig.~\ref{fig:cme1} presents the time-distance plot from Sun to Earth. To convert $\epsilon(t)$ to distance from the Sun $r(t)$ in Fig.~\ref{fig:cme1}, two simple methods have been put forward: the ``Point-P" and ``Fixed-$\Phi$" methods \citep{kah07, woo09}. Basically, Point-P assumes that the CME's LE is a spherical front centered on the Sun, which yields \citep{how07}
 \begin{equation}
 r(t)=d~ \textrm{sin}~ \epsilon(t)
 \end{equation}
  where $d$ is the distance of the observer from the Sun. Fixed-$\Phi$ assumes the CME to be point-like, propagating radially along a constant angle $\Phi'$ to the observer,
  \begin{equation}
 r(t)=\frac{d~ \textrm{sin} ~\epsilon}{\textrm{sin}~ (\epsilon+\Phi')}.
 \end{equation}

  The results are shown in Fig.~\ref{fig:cme1}. Linking the observations with the timing $t_s (t_c)$ when the CME LE (core) hits  STEREO-B located at 226 $R_{\odot}$, the angle $\Phi$ can be derived with $\Phi=-30 (-26)^{\circ}$. (Angle $\Phi$ is measured from Earth while $\Phi'$ is measured from an observer, negative means east of Earth).
  In Fig. \ref{fig:3D} the mean of the two directions is plotted. It is also seen in
  Fig.~\ref{fig:cme1} that until $r \approx 140 R_{\odot}$ or 0.62 AU, the Point-P and Fixed-$\Phi$ methods yield
  quite similar values for $r(t)$, with $r(t)$ being slightly lower for Point-P close to the Sun \citep[see also][]{woo09}.
  After this, the unphysical assumption of Point-P that the CME cannot be tracked farther than $d$ leads to a divergence from
  Fixed-$\Phi$ to smaller values of $r(t)$.


We now compare the directions obtained from the interplanetary techniques with direction--finding techniques
derived from coronagraph observations. Fig.~\ref{fig:cme1} (inset) shows the kinematics of a distinct feature along the CME
 leading edge, one that could be observed and measured from LASCO C2/C3 and COR2A images. As the feature is seen from
 different viewing angles, the de-projected propagation direction of the CME is derived by applying a geometrical triangulation method
 as described in \cite{tem09}.  The same feature has been tracked in the HI1A FoV (left yellow $\times$ symbols in Fig.~\ref{fig:hi}) which yields $\Phi=-45 \pm 5^{\circ} $.
 The applied technique is limited by the underlying assumptions (same CME feature distinguished
in used instruments, radial outward motion of CME). However, the results are consistent with
those derived by \cite{the09}, who use a forward modeling (FM) technique which
presumes a particular shape of the CME, and find  $\Phi=-37 \pm 10^{\circ}$. Table \ref{tab:cmedir} summarizes the various results of
 CME directions and de-projected velocities,
and the direction vectors are plotted in Fig.~\ref{fig:3D}. In summary, the methods yield roughly the same de--projected direction when compared to the MFR reconstruction, with a slight offset to the east by $\approx 10-15^{\circ}$ of the methods applied closer to the Sun. Concerning the latitudinal direction, TR and FM yielded results to within $\pm 5^{\circ}$
latitude of the solar equator, and the measurements for EF and KP were also taken within that range. This also means that STEREO-B encounters the apex or nose of the CME and not one of its ``legs''.

The CME velocities clearly show an acceleration between the coronagraphs and the heliospheric imagers as well as a consistency of the latter with the proton bulk velocities as measured in situ. From EF a velocity difference of $\Delta V\approx$ 60 km~s$^{-1}$ between the leading edge (401 km~s$^{-1}$) and the core (340 km~s$^{-1}$) is consistent with  $\Delta V= 49$ km~s$^{-1}$ between the MFR front and back boundary. These are values typical of ICMEs at 1 AU during solar minimum \citep[$\Delta V \approx$ 60-70 km s$^{-1}$:][]{jia06}; \citep[cf. 45 km s$^{-1}$ found by][]{bur82}. Note that both are larger than $\Delta V=19$ km~s$^{-1}$ between the averaged velocities of the two in situ density peaks which correspond to the white-light features.

\section{Conclusion}

\begin{figure}[h]
 \centerline{\includegraphics[width=20pc, bb=0 0 300 550]{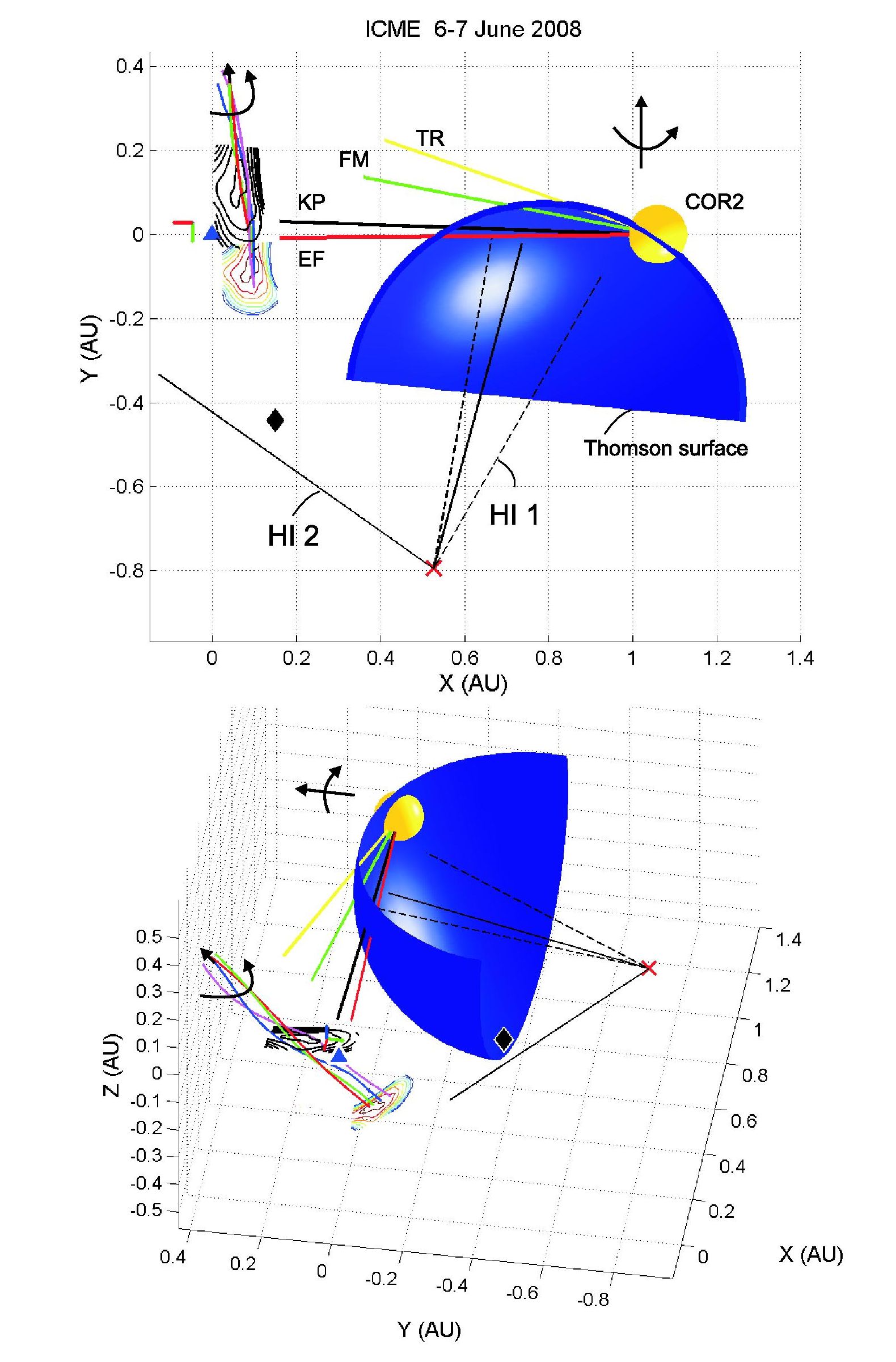}}
 \caption{3D plot of the local MFR reconstruction in the heliosphere (color contour) in a coordinate system centered on STEREO-B, seen from north looking down on the RT plane (close to the solar equatorial plane, top) and from another view point (bottom). For further details see text. This figure is also available as an mpeg animation in the electronic edition of the
{\it Astrophysical Journal}.}\label{fig:3D}
 \end{figure}

\begin{figure}[h]
\noindent\includegraphics[width=20pc, bb= 0 0 300 550 ]{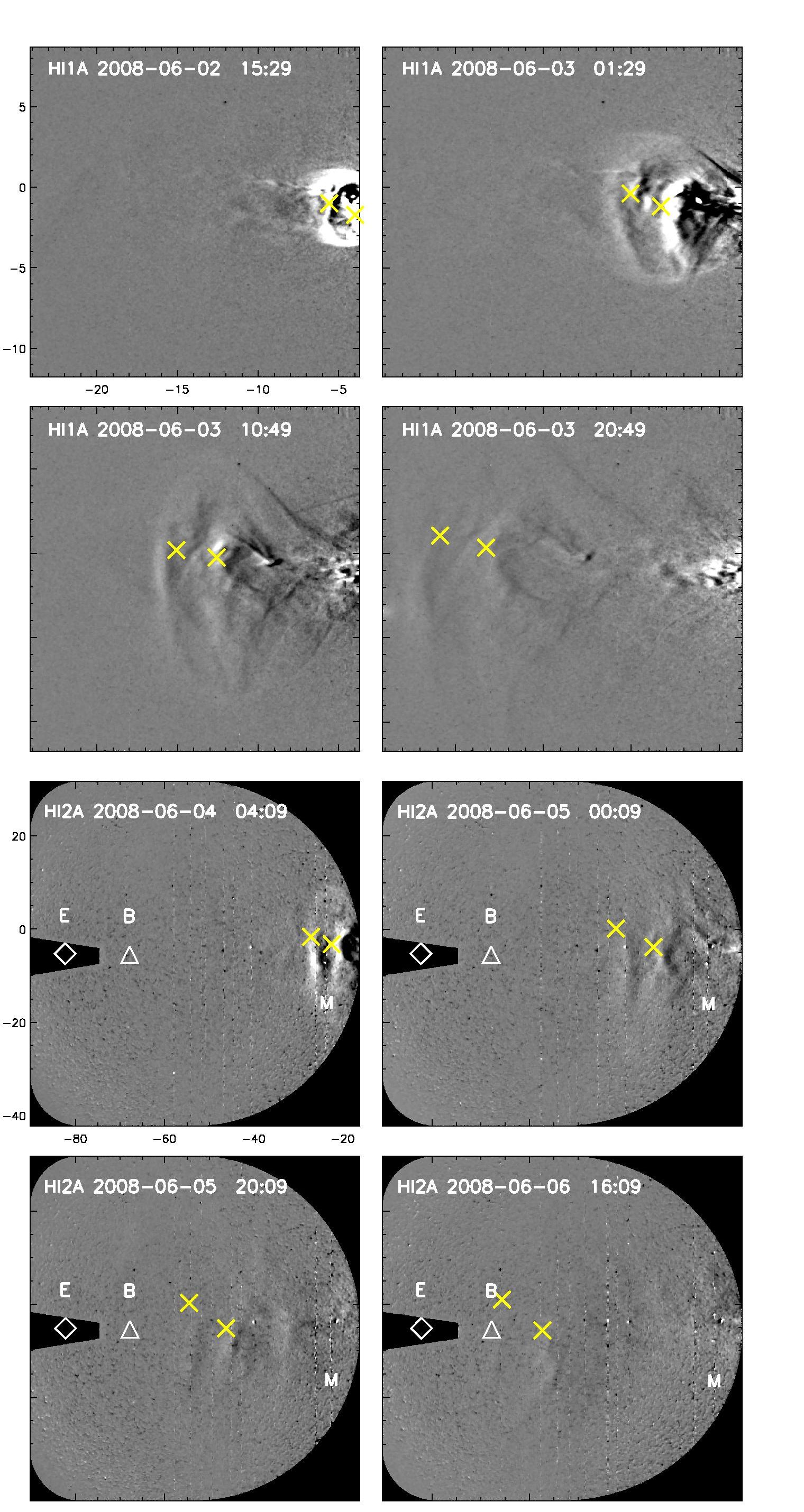}
 \caption{Evolution of the CME in STEREO-A HI1 (top 4 images) and HI2
 (bottom 4 images), cf. Fig.~\ref{fig:3D}. Earth (E), STEREO-B (B) are indicated as well as the elongation of Mercury. The tracked features of the CME leading edge and core for Fig.~\ref{fig:cme1} are given by yellow crosses.
 The last image was taken approximately at the shock arrival time at STEREO-B (15:35 UT). 
  Note the distinct appearance of the CME in both FoVs as an arc-like shape, i.e. a CME seen perpendicular to the axis of symmetry.
  This figure is also available as an mpeg
animation in the electronic edition of the
{\it Astrophysical Journal}.}
 \label{fig:hi}
\end{figure}

We described observations of a CME as it propagates from the Sun to 1 AU and detailed  the relationship to its modeled in-situ signatures. From this we found various interesting clues useful for forecasting the geoeffects of CMEs:

   1. How can the internal structure of the MFR be inferred from remote observations? The distinct arc-like morphology points to an
   inclined MFR axis and,  from the intensity evolution perpendicular to the ecliptic plane (northern part vanishes earlier than the southern), we estimated $\theta \approx \pm 45^{\circ}$. The MFR's axial and poloidal field directions are consistent with the global solar dipole field
   \citep[see the PFSS model in][]{rob09} and indicated by black arrows in Fig. \ref{fig:3D}: The poloidal field goes from south to north and
   the eruption comes from the southern hemisphere, which yields a right-handed MFR (as observed), and thus an axial field to the east
   (away from the observer, thus $\theta \approx +45^{\circ}$, $\varphi \approx 270^{\circ}$). The MFR's axial field
   at the Sun is likely to be close to the solar equator. This can be seen in an accompanying movie in \cite{rob09}, which shows a more compact CME shape closer to the Sun in accordance with the \cite{cre04} interpretation.  A slight subsequent clockwise rotation  (consistent with positive chirality for a twist-to-writhe conversion) of roughly $45^{\circ}$ would then match the MFR observations.

  2. Is the direction of the CME obtained from various direction-finding techniques consistent with the
observations at 1 AU? Yes, within the error bars  all methods we use give a consistent picture. There seems to be a minor offset to the west when techniques are applied closer to the Sun. Whether this is due to a bias of the different techniques or a slight non-radial CME propagation is the subject of further investigations.

  3. What is the magnetic flux content of this pre-existing magnetic flux rope on the Sun without reconnection signatures on the solar surface?
   We find that it is surprisingly high, comparable to events connected to C or even M-class flares \citep{qiu07}.
   Thus, this can be seen as evidence for a formation of the magnetic flux rope before the eruption for this particular event.


 The tools and concepts we developed allow a first comparison between 3D CME directions and modeled in-situ data of those ICMEs which contain magnetic flux ropes or magnetic clouds.
Modeling ICMEs is important in order to obtain good estimates of their orientations and impact parameters, and
where the spacecraft intersects the axis (the eastern/western part of the central MFR south/north).
Our results underline the need of having coronagraphs and instruments which image the heliosphere between the Sun and
the Earth from a vantage view-point away from Earth so as to enhance our ability to forecast the geo-effects of coronal mass ejections.

\acknowledgments
The authors are grateful to the referee for many helpful comments.
C. M\"ostl, C. Miklenic, A.V. and H.K.B. acknowledge the Austrian Science Foundation (\emph{Fonds zur F\"orderung der
wissenschaftlichen Forschung}) for support under project P20145-N16. A. B. Galvin is PI, and C.J.F. is a Co-I on STEREO/PLASTIC.
 This work is supported by NASA grants NAS5-00132, NNG06GD41G and NNX08AD11G.
  M. Temmer acknowledges project APART 11262 of the Austrian Academy of Sciences.
  We thank the STEREO/SECCHI teams for their open data policy.

\clearpage





\clearpage

\begin{table}
\begin{center}
\caption{CME directions and velocities from various techniques. Angle $\Phi$ is the longitude measured from Earth (negative means eastward).  $R_{\odot}$ gives
the distance from the Sun of the observations used for the respective technique and V the velocity for this distance range.
 \label{tab:cmedir} }

\begin{tabular}{ccccc}
\tableline\tableline
Technique & Instruments & $R_{\odot}$&$\Phi$, deg &  V,  km s$^{-1}$ \\
\tableline
Triangulation (TR) & C2/C3 COR2A & 3-14 & -45 $\pm 5$ & 235 \\ 
Forward modeling (FM)\tablenotemark{a} & COR2A/COR2B & 2-15 &-37 $\pm 10$ & 265\\
Elongation fitting (EF)\tablenotemark{b} & HI1A/HI2A & 15-226 & -26 (-25)  $\pm 3$  & 401 (340) $\pm$ 15 \\
Kinematic Fixed-$\Phi$  (KP)\tablenotemark{b} &  HI1A/HI2A & 80-226 &-30 (-26) $\pm 5$ & 440 $\pm$ 120\tablenotemark{c} \\
In-situ modeling &   IMPACT/PLASTIC  & 226 & $-24$ to $-39$   & 403 / 392 / 384\tablenotemark{d} \\
\tableline
\end{tabular}
\end{center}
 \tablenotetext{a}{taken from \cite{the09}.}
 \tablenotetext{b}{Results for the CME core are in brackets after those for the leading edge.}
 \tablenotetext{c}{Here $V$ is the median velocity from 80-226 $R_{\odot}$ in a $V(r)$ plot. In this range $V(r)$ is roughly constant.}
 \tablenotetext{d}{The in-situ $V$ are means over the sheath region, the deHoffmann-Teller velocity of the MFR region, and a mean over the second density peak, respectively.}
\end{table}

\clearpage

\clearpage

\clearpage

\clearpage

\clearpage

\clearpage






\end{document}